\documentclass[AMA,Times1COL]{WileyNJDv5} 
\usepackage{tcolorbox}

\articletype{Article Type}%

\received{Date Month Year}
\revised{Date Month Year}
\accepted{Date Month Year}
\journal{Journal}
\volume{00}
\copyyear{2023}
\startpage{1}

\begin{document}
\title{A Reliable Self-Organized Distributed Complex Network for Communication of Smart Agents}

\author[1]{Mehdi Bakhshipoor}

\author[1]{Yousef Azizi}

\author[1,2]{Seyed Ehsan Nedaaee Oskoee}

\authormark{OSKOEE \textsc{et al.}}
\titlemark{A Reliable Self-Organized Distributed Complex Network for Communication of Smart Agents}

\address[1]{\orgdiv{Department of Physics}, \orgname{Institute for Advanced Studies in Basic Sciences (IASBS)}, \orgaddress{\state{Zanjan 45137-66731}, \country{Iran}}}

\address[2]{\orgdiv{Research Center for Basic Sciences and Modern Technologies (RBST)}, \orgname{Institute for Advanced Studies in Basic Sciences (IASBS)}, \orgaddress{\state{Zanjan 45137-66731}, \country{Iran}}}

\corres{Corresponding author Seyed Ehsan Nedaaee Oskoee, This is sample corresponding address. \email{nedaaee@iasbs.ac.ir}}

\abstract[Abstract]{
	Collaboration among distributed agents is fundamental to many complex systems, particularly in communication networks where connectivity must be maintained under energy constraints. In this study, we utilize intelligent agents (nodes) trained through reinforcement learning techniques to establish connections with their neighbors, ultimately leading to the emergence of a large-scale communication cluster. Notably, there is no centralized administrator; instead, agents must adjust their connections based on information obtained from local observations. The connection strategy is formulated using a physical Hamiltonian, thereby categorizing this intelligent system under the paradigm of "Physics-Guided Machine Learning" \cite{Seyyedi_Bohlouli_Oskoee_2023}. 

	Agents are trained via a Deep Q-Network using local observations to minimize changes in the Hamiltonian, enabling adaptive decision-making in dynamic environments. Simulation results demonstrate that the proposed collaborative strategy forms robust large-scale communication clusters while reducing transmission energy compared to baseline approaches. The network maintains high connectivity under agent mobility, density variations, node failures, and environmental obstacles, highlighting strong adaptability and resilience. These findings indicate that physics-guided reinforcement learning provides an effective mechanism for distributed topology optimization in emerging IoT and vehicular communication networks.	
}

\maketitle

\section{Introduction}

Complex systems in various scientific fields are recognized as patterns of interactions among components in irregular and self-organizing environments. These systems typically exhibit nonlinear and unpredictable behavior resulting from a phenomenon known as emergence. Emergence refers to the appearance of macroscopic features in a system through simple, local interactions among its components \cite{Hanel_2018}. In this study, we investigate the emergence of complex networks and how intelligent clusters are formed and reinforced through collaboration among the components.

In many natural systems, interactions between individual elements lead to the emergence of collective behaviors that enhance the overall performance of the system. In complex networks, the distributed nature of these interactions is a key feature that offers numerous advantages. Decentralized or distributed networks, unlike centralized systems, exhibit greater flexibility and resilience, as disruptions in one part of the system have limited effects on overall performance. This characteristic also enables dynamic adaptation to environmental changes and varying loads, resulting in higher efficiency and reduced security vulnerabilities \cite{Hanel_2018}.

One type of distributed complex network where emergence can be observed is the Internet of Things (IoT). The term "Internet of Things" refers to a system of interconnected entities that can interact and collaborate to achieve common goals through specific protocols \cite{Seyyedi_Oskoee_2023}. These networks serve as an excellent example for studying collaboration in complex networks due to their numerous applications in everyday life. However, how do these networks operate, and how should we model them?

\begin{figure}[t]
	\centering
	\includegraphics[width=0.96\linewidth]{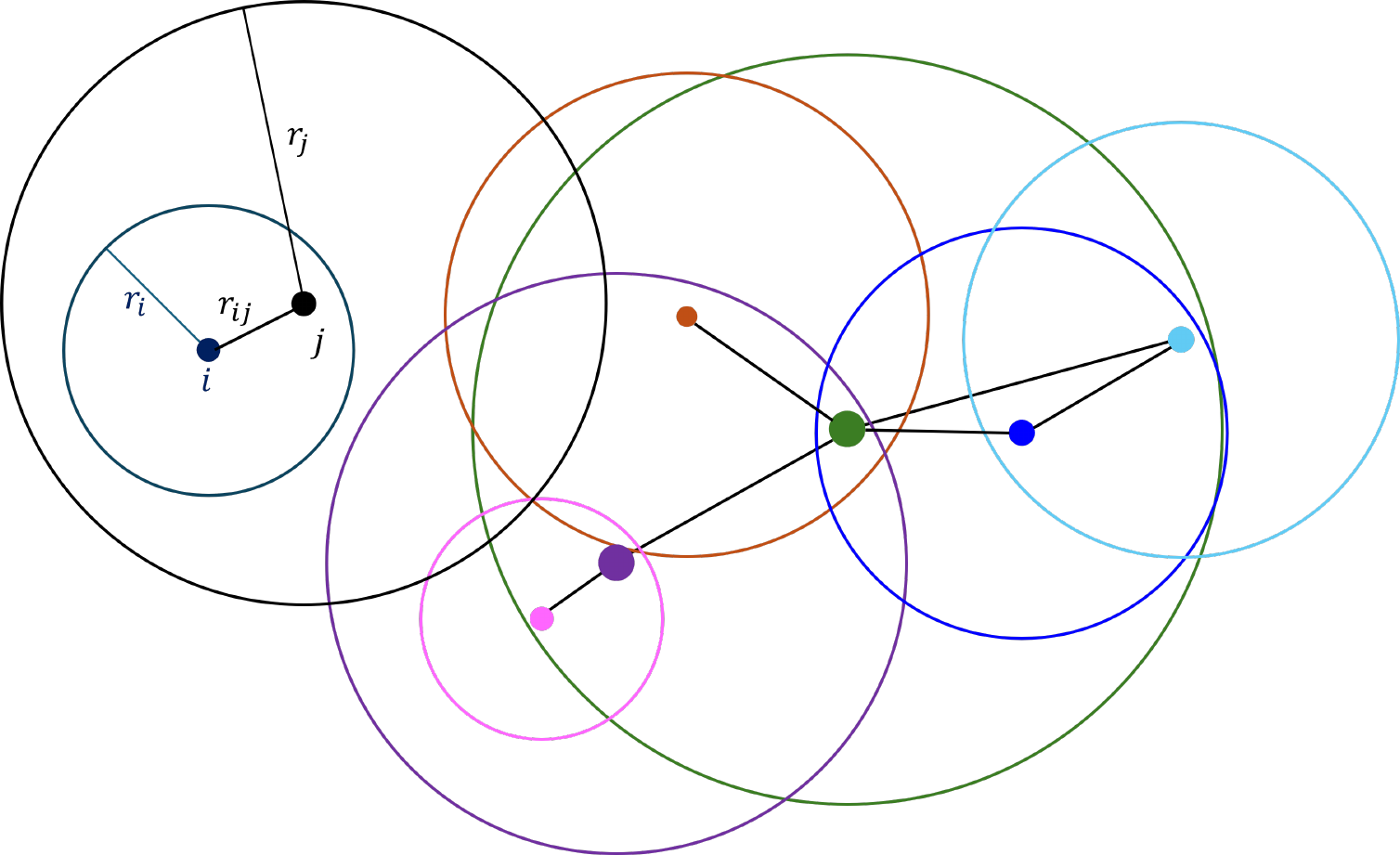}
	\caption{The spatial distance between members and the translation radius of each member determines the connection between elements. If two network members are located within the translational radius of each other, these two members are connected.}
	\label{fig: Graph with Transmission Range}
\end{figure}

For information to be transmitted within the Internet of Things (IoT) network, each member must function as both a sender and a receiver. Data is sent from one member using electromagnetic waves and received by other members. This establishes a form of connectivity within the network; if two members are within each other's transmission range, they are considered connected; otherwise, they remain unaware of each other (Fig. \ref{fig: Graph with Transmission Range}). Therefore, the likelihood of communication between devices, and consequently the overall connectivity of the network, increases with the expansion of each member's transmission range. However, this proposed solution for routing and overall network connectivity is not without its drawbacks. Most devices rely on batteries as their power source, which have limited capacity. Increasing the transmission range leads to higher energy consumption, ultimately reducing the longevity of network connections \cite{Cardei_Du_2005}. A longer transmission range can also result in excessive interference and communication disruptions \cite{Zengen_2014}.

As mentioned, point-to-point connections, one traditional method of communication between components, is not very efficient. Wireless IoT networks can collectively decide to adopt a not-so-large radius and transmit information by creating an appropriate path \cite{Cardei_Wu_2004}. Given that network topology significantly impacts its performance, it seems that improving network performance is possible through conscious management of its topology. The main idea behind controlling the network topology is maintaining connectivity while minimizing energy costs and interference \cite{Chiwewe_Hancke_2011}. This goal can be achieved by minimizing the transmission range of each member while still maintaining connections with its nearest neighbors.

Existing research on topology control in wireless and IoT networks has extensively explored resilience, connectivity guarantees, and architectural design from algorithmic or protocol-oriented perspectives. For instance, comprehensive surveys emphasize maintaining k-connectivity and fault tolerance through deployment-stage strategies and routing mechanisms, while identifying the inherent combinatorial complexity and open challenges in topology optimization. Such works highlight that topology control problems are frequently NP-hard and constrained by unreliable links and limited energy resources, motivating heuristic or stage-based solutions rather than adaptive self-organizing mechanisms \cite{huang_2015}. Likewise, broader studies on wireless sensor network architectures describe energy efficiency, layered protocol management, and security concerns as central design factors but typically rely on predefined structural assumptions rather than emergent behavioral adaptation \cite{mahmud_2013}.

More recent approaches have incorporated machine learning and reinforcement learning for distributed decision-making in wireless environments. Deep reinforcement learning has been applied to topology optimization or resource allocation, often improving adaptability compared to heuristic methods; however, many of these frameworks depend on search-guided optimization, centralized training procedures, or problem-specific heuristics that limit interpretability and physical grounding of the objective function \cite{Meng_2019, lu_2021, pi_2025}. Even fully distributed multi-agent reinforcement learning schemes generally address transmission scheduling or resource allocation without embedding domain-inspired structural constraints into the learning objective itself \cite{farquhar_2022}. Consequently, a gap remains between physics-informed modeling of network structure and adaptive decentralized learning strategies. The present work addresses this gap by integrating a Hamiltonian-based, formulation—capturing geometric and topological characteristics, with continuous reinforcement-driven adaptation, enabling physically interpretable, self-organizing topology control under dynamic conditions.

A physics-based approach is proposed here as a foundation for controlling the topology of distributed networks. This method utilizes Hamiltonian mechanics from physics, taking into account the intrinsic characteristics of Networks to implement an optimal solution. By considering the studied network as a complex physical system and analyzing its structural features and existing connections, a Hamiltonian function is created for the system \cite{Marion_2004}.

In this context, the Hamiltonian function serves as a cost function, intending to find the Hamiltonian minimum or equilibrium point. This approach effectively addresses the problem of adjusting the minimum transmission radius necessary to establish an optimal topology, which corresponds to the equilibrium state of the system.

In dynamic large-scale wireless networks, maintaining reliable connectivity while ensuring scalability and energy efficiency remains a fundamental challenge, particularly in scenarios such as mMTC and vehicular ad hoc networks (VANETs). Existing studies have explored adaptive clustering, mobility-aware topology management, and traffic-aware routing strategies to address these issues. For instance, recent works propose dynamic cluster head management mechanisms for energy-efficient IoT operation, as well as traffic-aware routing protocols that incorporate environmental factors such as mobility patterns and traffic signals to improve connectivity stability \cite{ETAR-2015, SPARC-2025, UAV-Caching-6G-RL-2026}.
However, these approaches typically rely on predefined heuristics or centralized coordination, whereas the proposed framework leverages fully decentralized physics-guided reinforcement learning for adaptive topology formation.

The primary focus of this text is to present a model for the topology of decentralized or distributed complex networks in both two and three dimensions using the Hamiltonian function of the network. A key feature of this approach is its ability to self-organize the network topology intelligently, transforming it from a completely disconnected configuration to an almost connected one (with connectivity greater than 90\%). This connectivity is resilient to small-scale changes within the network and can quickly restore the original connectivity following larger-scale events, such as attacks or the loss of a significant percentage of members. Our approach proposes a balanced description of the system to achieve optimal connectivity along with reasonable energy consumption. Using adjustable parameters in the Hamiltonian, the system can be adjusted to operate in the desired state in terms of energy and connectivity.

(In a previous article from our research group, a similar problem was investigated using a machine learning approach guided by physics, employing a simple MLP, which showed promising results \cite{Seyyedi_Bohlouli_Oskoee_2024}. In this work, we have utilized reinforcement learning more extensively, keeping the agents continuously learning, which has significantly enhanced the robustness and adaptability of the network.)

\section{Proposed Model}

\subsection{Hamiltonian Cost Function}
In its simplest definition, a network is a collection of nodes connected by edges \cite{Newman_2018}. This definition allows us to utilize graph theory as an abstract mathematical concept to describe the structure and behavior of networks. In the context of the Internet of Things (IoT), the nodes represent the devices within the network, each located at a specific geographic position, while the edges indicate the existence of communication between these devices. When an edge exists, it means that the two nodes can "see" each other or, in other words, are within each other's transmission range, allowing for information exchange between them. This information transfer occurs bidirectionally, indicating that the network under consideration is undirected; mathematically, this is interpreted as the adjacency matrix being symmetric.

It has been established that each member's location represents a node's position within the network. Therefore, the network can be considered a Euclidean spatial graph, where the Euclidean distance defines the connectivity metric. This means that if the spatial distance between two members is less than their transmission range, a connection will be established between them. As illustrated in Fig. \ref{fig: Graph with Transmission Range}, the network can be analyzed in a two-dimensional space, and it can then be easily extended to a three-dimensional representation \cite{Seyyedi_Oskoee_2023}.

To build the network, we use the Hamiltonian equation as a cost function, which includes the topological characteristics of the network and the way of communicating between the components in its definition. This Hamiltonian is defined as
\begin{equation}
	H = \sum_{i} H_i = \sum_{i} \left[ \alpha_1 k_i^2 + \alpha_2 k_i^3 + \alpha_3 r_i^2 + \alpha_4 \sum_{\substack{(j \ne i)}} \frac{A_{ij}}{r_{ij}} \right].
	\label{eq: Hamiltonian}
\end{equation}
In this equation, $i$ represents each node and $k$ indicates the degree of the node or the number of links connected to it. $r$ is the transmission radius of the node, $r_{ij}$  is the distance between two arbitrary nodes $i$ and $j$, and A represents the adjacency matrix of the network. In this matrix, if a link exists between two nodes, the corresponding row and column for those nodes will symmetrically equal 1; otherwise, their value will remain 0. Also, the coefficients $\alpha_{1}$ to $\alpha_{4}$ are arbitrary constants that may take on different values depending on the specific conditions of the problem and the number of nodes \cite{Berg_Lassig_2002}. The optimal coefficient values were calculated in our previous work \cite{Seyyedi_Oskoee_2023}, and here we will simply use those obtained values. We do not treat them again as separate hyperparameters for which we intend to perform additional computation or simulation.

The first two Hamiltonian equations related to the topological distribution of the network were proposed by Johannes Berg and Michael Lässig. The first sentence indicates the number of paths of length two within the network. To form short paths of length two, nodes strive to connect to one or only a few other nodes; thus, the Hamiltonian rewards the formation of hubs within the network. In other words, the first Hamiltonian sentence drives the configuration toward the most compact and star-like arrangement possible. The second sentence prevents the Hamiltonian from collapsing into a purely star-shaped configuration or a single large hub by introducing a penalty. This term, often referred to as regularization, includes a positive coefficient that curbs excessive increases in node degree and promotes more connections involving three nodes (triangular connections), which in turn leads to a more distributed network \cite{Berg_Lassig_2002}.

The second two terms are called geometric terms. The third term of the equation pertains to the transmission power of the nodes. From electrodynamics, we know that the energy of electromagnetic waves decreases with distance from the source according to the inverse square law. As $r$ increases, the \textit{radiation} (energy per unit area) diminishes with the square of the distance from the source \cite{Griffiths_2013}. Therefore, to achieve a longer transmission radius, more initial energy from the source is required, which in turn leads to increased battery consumption. Thus, to optimize energy usage, the radius should be kept as small as possible, although this may result in the formation of some sub-clustered networks that are not connected to the main network. The presence of the fourth term is intended to prevent this situation. This term rewards the Hamiltonian by allowing for a few long links within the network, ensuring that even these small isolated clusters become connected to the main component.

The total Hamiltonian is the sum of the Hamiltonians of each member, which aligns with the definition of this physical function. The Hamiltonian of the system, considering the coefficients $\alpha_{1}$ to $\alpha_{4}$, seeks to find the equilibrium state. To achieve this equilibrium, the Hamiltonian reduces the cost and energy consumption of the network by adjusting the transmission ranges of the nodes, while also striving to maximize connectivity. These two parameters interact in such a way that they are influenced by environmental conditions such as the number of nodes per unit volume (node distribution density), which converge to a stable equilibrium point corresponding to the minimum Hamiltonian.

Each of the coefficients in the Hamiltonian function influences the emergent topology in different ways, effectively determining the sensitivity of each stated proposition. For instance, increasing $\alpha_{2}$ relative to $\alpha_{1}$ suppresses star-shaped configurations and leads to a higher clustering coefficient and greater degree homogeneity, whereas lower values of $\alpha_{2}$ allow dominant hubs to emerge. As another example, the coefficient $\alpha_{3}$ directly controls the network's average power consumption. Larger values of $\alpha_{3}$ lead to smaller transmission ranges and may cause network fragmentation in low-density regimes, while smaller $\alpha_{3}$ values promote global connectivity at the cost of reduced energy efficiency. Finally, the coefficient $\alpha_{4}$, which controls the sensitivity of long-range links, exhibits the following behavior: excessively large values of $\alpha_{4}$ increase energy consumption by encouraging long-range connections.

Simulations demonstrate that across a wide parameter range around the chosen values, the network exhibits stable connectivity patterns, proving that the proposed approach is robust to gentle variations in the coefficients, rather than relying on fine-tuned parameter selections. A detailed numerical sensitivity analysis of these coefficients has been presented in our group's previous paper \cite{Seyyedi_Oskoee_2023}, so we have refrained from providing further explanations here. However, to demonstrate the validity of the described qualitative behavior using a method different from our previous work, we have, for example, also subjected the coefficient $\alpha_{4}$ to numerical investigation in part \ref{sec: hyperparameters Effect} of the Results section.

\subsection{Reinforcement Learning Algorithm}
Reinforcement learning is one of the most exciting and oldest areas of machine learning, originating in the 1950s and producing many interesting applications over the years, particularly in games and control systems. The main difference between this method and the well-known supervised learning approach is that in supervised learning, the algorithm learns from labeled samples, aiming to map inputs to predefined outputs based on the provided training data. In contrast, reinforcement learning operates in an environment where the algorithm learns through trial and error \cite{Geron_2019}. While supervised learning involves guidance from a knowledgeable "teacher," reinforcement learning removes this guiding figure, allowing the algorithm to learn by experiencing different states and making decisions based on its own experiences.

In reinforcement learning, the \textbf{agent} first observes its surrounding \textbf{environment} and takes \textbf{actions} based on that observation, receiving \textbf{rewards} in return. The goal is for the agent to learn to act in a way that maximizes the rewards it receives over time. The humanistic interpretation of this process is that positive rewards are perceived as pleasure. In contrast, negative rewards are viewed as pain (the term "reward" can be somewhat misleading in this context). In summary, the agent performs actions in the environment and learns through trial and error to maximize its pleasure and minimize its pain \cite{Sutton_1998}.

In the mid-twentieth century, Richard Bellman introduced the concept of Markov Decision Processes (MDPs) based on Markov chains and reward-punishment systems \cite{Sutton_1998}. In an MDP, an agent can choose from several possible actions at each state, and the probability of transitioning to the chosen action depends on that choice. Furthermore, some transitions return a reward (either positive or negative), and the agent's goal is to find a policy that maximizes rewards over time \cite{Geron_2019}. Bellman devised a method to estimate the value of each state by introducing the recursive Bellman optimality equation \cite{Sutton_1998}.

The Bellman optimality equation is generally valid and leads us to the optimal value of each state; however, in real-world problems, implementing this algorithm is often not feasible. In the Bellman optimality equation, due to the presence of the transition probability term, utilizing such model-based algorithms becomes practically impossible if the environmental model is not fully known. Therefore, model-free algorithms will be beneficial \cite{Sutton_1998}.

The temporal difference algorithm is an environment-independent method that is useful for systems that are continuously interacting. In this approach, the system evaluates its decision after each action and updates itself based on the received reward. The equation used to obtain the action-value function in the temporal difference algorithm is expressed as:
\begin{equation}
	Q(s,a) \leftarrow Q(s,a) + \eta \left( \left[ r + \gamma \max_{a^{'}} Q(s^{'},a^{'})  \right] - Q(s,a)  \right).
	\label{eq: Q-Learning}
\end{equation}

In this equation, $\eta$ represents the learning rate, $\gamma$ is the discount factor, a parameter that weakens the effect of rewards in the distant future, and $r$ is the reward for transitioning from state $s$ to state $s'$. The term inside the brackets represents the target value of the temporal difference, while the entire expression within parentheses is referred to as the temporal difference error. Here, instead of using the state value, we utilize the action-value function $Q(s, a)$ for action $a$ in state $s$ \cite{Geron_2019, Sutton_1998}. The state value can be viewed as a generalized form of $Q(s, a)$. It can be observed that in this method, the target value is of the same nature as the state values; that is, it does not have direct knowledge of the actual final target value but intelligently estimates it for the next step. The system then attempts to approach this target using a method similar to gradient descent \cite{Sutton_1998}.

\section{Simulation Details}
Now, we will implement an algorithm to train a simple deep neural network model using the previously stated relationship for the temporal difference and the associated error. In complex environments with large state spaces, storing and updating all action-value pairs $Q(s, a)$ can be very resource-intensive. Therefore, we will turn to deep neural networks and utilize the more advanced and integrated approach known as Deep Q-Network (DQN).

The implemented algorithm for the stated method is shown in procedural form below.

\begin{table}[ht]
	
	\begin{tcolorbox}[title={\raggedright \hspace*{-0.3cm} Our DQN method training}]
		\raggedright
		
		\hspace*{-0.3cm} Initial: Parameters $\eta = (0,1]$, $\gamma = [0,1]$ and $\varepsilon > 0$ \\
		\hspace*{0.55cm} Neural Network (with random variables) \\
		\hspace*{0.55cm} Initial state of agents $s$ \\‌‌ \\
		
		\hspace*{-0.3cm} Loop for each step: \\
		\hspace*{0.2cm} Update $\varepsilon$ based on step
		
		\hspace*{0.2cm} Choose $a$ from $s$, using NN ($\varepsilon$ greedy) \\
		\hspace*{0.2cm} Take action $a$, and observe $s'$, $r (=-\Delta H)$ \\     
		\hspace*{0.2cm} Update neural network weights as:
		
		\hspace*{0.5cm} $Q^{*}(s',a') = \max_{a'} \left( NN.predict(s', a') \right) $ \\
		\hspace*{0.5cm} $Target = r + \gamma \times Q^{*}(s',a')$ \\
		\hspace*{0.5cm} Calculating the error with the $Target$ value \\
		\hspace*{0.5cm} Optimization of neural network weights
		
		\hspace*{0.2cm} $s' \rightarrow s$
		
	\end{tcolorbox}
	
	\label{tab: Q-Learning with NN Algorithm}
\end{table}

The value of $\varepsilon$ mentioned here represents the degree of greediness of the algorithm in making random decisions versus decisions based on previous learning. Initially, this value is set close to 1 to encourage more random actions at the beginning of the decision-making process, allowing the agents to effectively explore various states. Over time, the decision-making becomes increasingly greedy, and the agents begin to act according to their prior learnings, guiding them toward making more optimal decisions.

\begin{figure}[b]
	\centering
	\includegraphics[width=\linewidth]{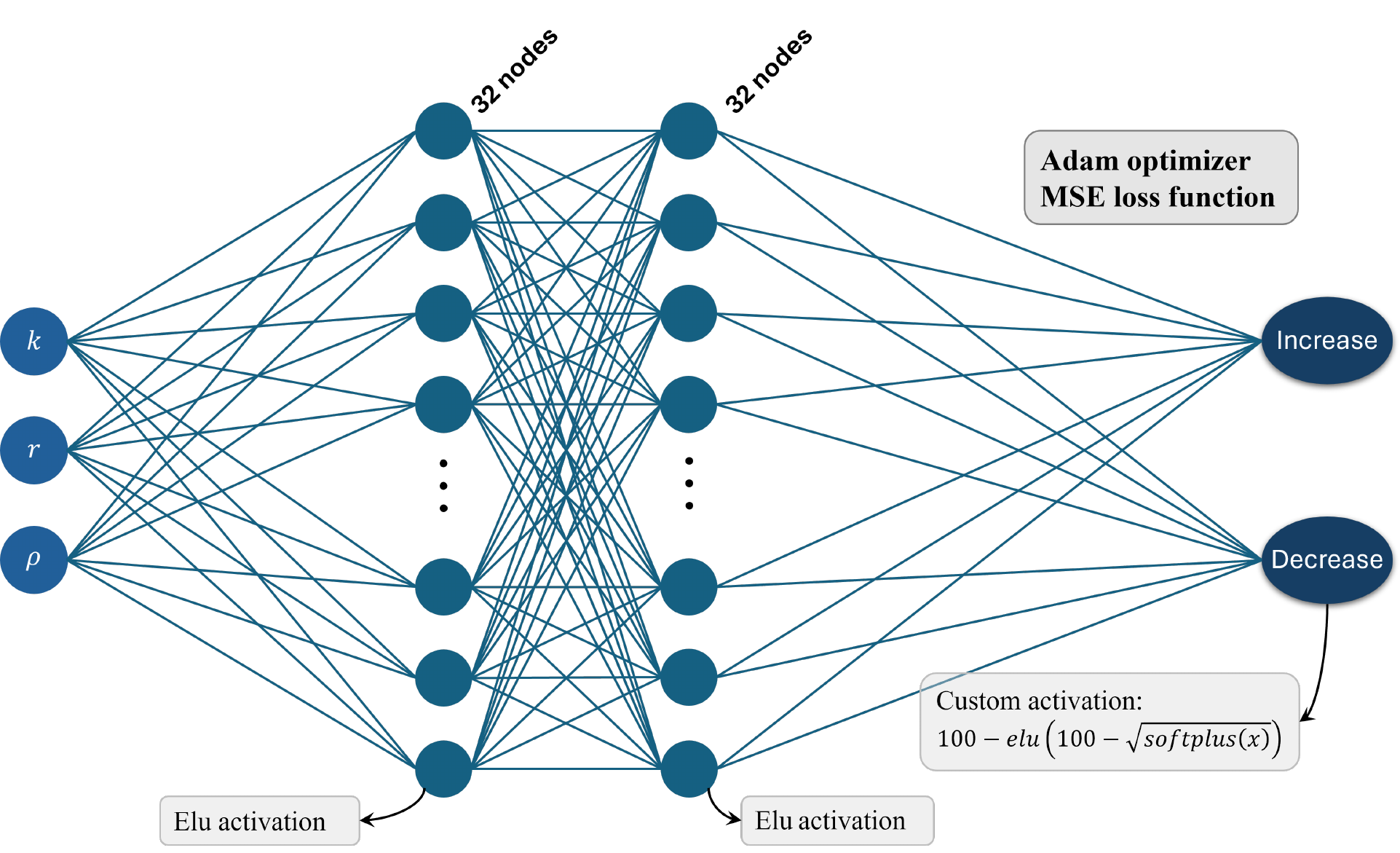}
	\caption{The neural network diagram used features as (a layer with three nodes) input, two hidden layers with 32 nodes each, and a final output layer consisting of two neurons to determine whether to increase or decrease the agent's radius. The model's loss function is the mean squared error, and the optimizer employed is the Adam optimizer with an appropriate learning rate. Additionally, the activation function for the two initial hidden layers is the Exponential Linear Unit (ELU), while the output layer utilizes a generalized version of the ELU function.}
	\label{fig: Neural Network Model}
\end{figure}

Updating the neural network weights corresponds to the training phase, essentially a generalization of Eq. (\ref{eq: Q-Learning}) for deep networks. Updating the weights in the correct direction allows the model to find optimal weights after training, enabling it to make the best decisions in various conditions based on the value of those actions.

The neural network is a simple deep network, as shown in Fig. \ref{fig: Neural Network Model}. The network receives the state of the agent (the density of surrounding agents, its current transmission radius, and the number of agents it is connected to), and ultimately, it determines whether the agent should increase or decrease its radius. The output value from each of the increase or decrease neurons indicates which action the agent should take; the action corresponding to the neuron with the higher output value is the one the agent should perform. By selecting this action, the Hamiltonian changes, and the magnitude of this change ($-\Delta H$) represents the reward received by the agent for taking that action.

\begin{figure*}[t]
	\includegraphics[width=\textwidth]{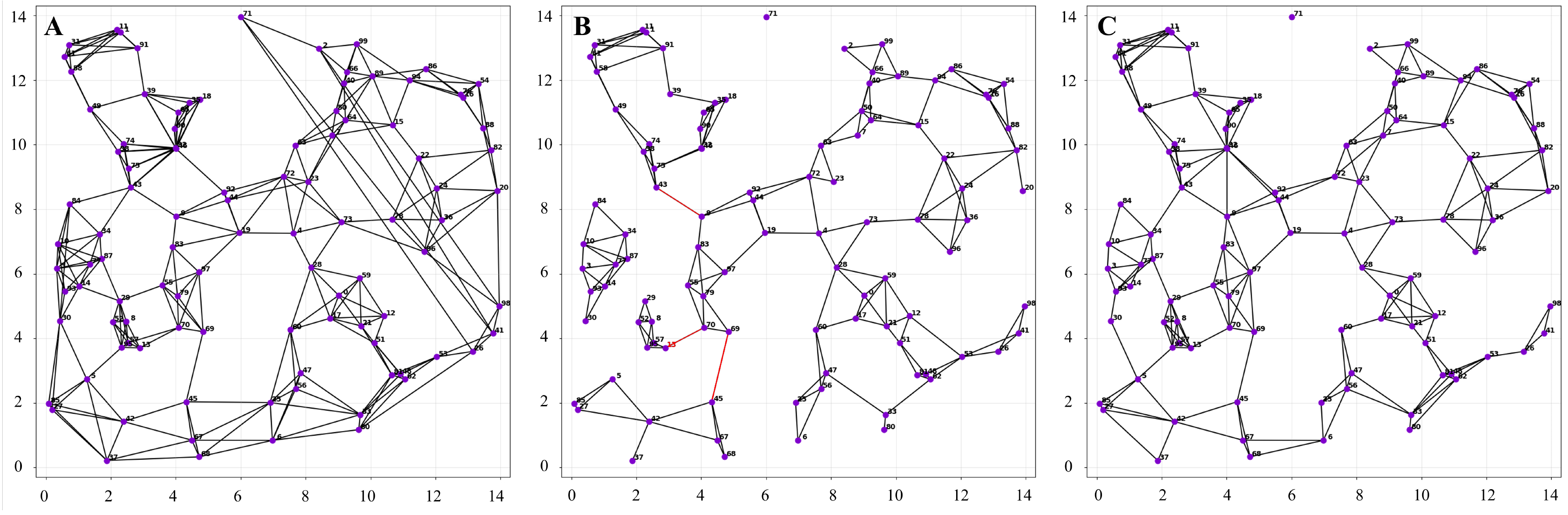}
	\caption{A snapshot illustrating the three examined approaches is presented. A is the classical minimum k-degree approach, where agents choose a large radius to achieve complete connectivity, consuming significant energy. B corresponds to the neural network approach, which demonstrates the agent clustering. There are some critical (red) links in this approach, which, if removed them cause partitioning of the giant cluster. C is the same as B, but including agent partial collaboration by the request sending process. In contrast with B, the chance of red link establishment is lower in C, resulting in a more robust giant cluster.}
	\label{fig: Environments}
\end{figure*}

To improve training stability, a lightweight experience replay mechanism is employed, where each agent maintains a local buffer of the most recent interactions (with a maximum size of 48). During training, mini-batches (of size 16) are sampled uniformly at random from this buffer, which helps reduce temporal correlations between consecutive updates despite the fully decentralized setting.

The simulation consists of two phases: the first phase is the initial training stage, which we have already outlined in the algorithm, and the second phase occurs after the model has been trained. The training phase aims to obtain optimal weights for the neural network (where optimality is defined as the point at which the total Hamiltonian of the network is minimized) and ultimately stores these weights. 

In the decision-making phase, each agent will have a dedicated neural network, with the initial weights being those learned during the previous training stage. The agents receive this neural network and continuously optimize their weights based on the new environmental conditions they encounter, effectively continuing the learning process. This secondary training occurs every few steps and is conducted with a lower learning rate to prevent noise from affecting learning. The graphs presented in Section 4 are all related to this fine-tuning training phase.

Also, exploration is only used during the initial training phase (an $\epsilon$-greedy policy), and is gradually reduced as learning progresses. After convergence, agents rely entirely on the learned policy without random exploration. We observed that variations in the exploration schedule primarily affect convergence speed rather than the final network configuration, indicating that the overall system behavior is not highly sensitive to the precise choice of $\epsilon$ scheduling.

In the following, four strategies are introduced, and the results are analyzed accordingly. The first and second strategies represent classical approaches, illustrating a simple and straightforward cases whose main purpose is to provide a reference for comparing other strategies and highlighting their advantages and applications. The third strategy involves intelligent agents that make optimal decisions using the neural network assigned to them. Finally, the fourth strategy, in addition to employing intelligent agents (i.e., decision-making based on the neural network), incorporates a mechanism for sending connection requests among agents in the network to achieve more optimal connectivity.

\begin{enumerate}
	\item \textbf{Minimum k-Degree strategy:} In this approach, one of the fundamental and classical algorithms for self‑organizing networks, decisions follow a random process that either increases the radius or leaves it unchanged. This process continues until each agent becomes connected to at least k other agents, ensuring that all agents belong to a single large connected component.
	\item \textbf{k‑Nearest Neighbor strategy:} This is another classical algorithm for self‑organizing networks, in which agents strive to establish connections specifically with their k nearest neighbors. In this case, agents gradually increase their communication radius until they are connected to k other agents, after which the radius remains fixed. If the number of connections exceeds k, the agent begins reducing its radius until it once again reaches the desired degree k.
	\item \textbf{Smart strategy:} Here, agents are initialized with pre-stored neural network weights, and decision-making is carried out based on this neural network. By observing their surrounding environment, agents determine their current state and decide whether to increase or decrease their transmission radius. In this way, each agent, without any knowledge of the decisions made by other agents, selects the best decision for itself.
	\item \textbf{Smart cooperative strategy:} This strategy is similar to the second strategy, in which each agent makes its decisions based on a neural network, with the difference that here we also observe a form of request exchange among agents. In this case, agents that, from their perspective, have not yet reached an optimal state, or are still connected to only a few other agents relative to the surrounding density, send their location information to others. Any other agent with a small transmission radius but within the range of the requesting agent receives this information. The receiving agent then calculates whether accepting the request (i.e., increasing its radius to cover the distance between the two agents) would result in a reduction of its Hamiltonian.
	
	The difference in Hamiltonian can be easily computed for the receiving agent (index $i$ indicates the receiver and index $j$ indicates the sender) using the following relationship.
	\begin{equation}
		\begin{aligned}
			\Delta H_{i} = \alpha_{1} \left( (k_{i}+1)^{2} -k_{i}^{2} \right) + \alpha_{2} \left( (k_{i}+1)^{3} -k_{i}^{3} \right) 
			+ \alpha_{3} \left( r_{ij}^{2}-r_{i}^{2} \right) + \alpha_{4} \left( \frac{1}{r_{ij}} \right).
		\end{aligned}
		\label{eq: Delta_H Request}
	\end{equation}
	It is also important to note that this calculated Hamiltonian difference considers only the scenario where the receiving agent connects solely to the requesting agent; there may be several other agents present that it could connect to or request them to connect. This situation fosters a type of collaboration among the agents, as the reduction in the Hamiltonian for the receiving agent also leads to a decrease in the Hamiltonian for the requesting agent and induces changes in the agents situated between them.
\end{enumerate}

\section{\label{sec: Result}Result and Discussion}

In this section, we describe the results of the Hamiltonian approach using DQN for network topology control. The parameters considered for this simulation are explained below. The number of agents used in this simulation (both during the learning and model testing phases) is $N=100$. The space designated for the placement of agents is a square-shaped area that accommodates varying densities of agents per unit area, with the range of $\rho = [1, 0.01]$ units (all units are arbitrary). The coefficients $\alpha_{1}$ to $\alpha_{4}$ will vary depending on the system length, and these coefficients have been adopted from our previous work \cite{Seyyedi_Oskoee_2023}. The discount factor is set to $\gamma=0.98$, and the learning rate is chosen based on the conditions to ensure that the agents and the system move toward a global minimum. To ensure robustness, several different models have been tested in the simulations, with the learning rate range specified as $\eta = \left[ 10^{-5},10^{-3} \right]$. Another parameter in this problem is the greedy policy coefficient $\varepsilon$. This value changes over time and is calculated according to the following relationship
\begin{equation}
	\varepsilon = \max \left( 1-\frac{t}{T_{max} / 2} , 0 \right),
	\label{eq: e-greedy}
\end{equation}
where $t$ represents the current time step, and $T_{max}$ is the total number of simulation steps.

The initial radius assigned to each agent is set to 1 unit. The change in transition radius of each agent according to the output value of the DNN model at each step is given by $\Delta r = \pm 1/4 \sqrt{L^{2}/N}$ units multiplied by a random value between 0 and 1. Adding this random component ensures that the changes are not always discrete, allowing the agents to experience values between these discrete states. To better display the results, the total number of steps considered for each simulation run is set to $T_{max}=1000$ steps.

Empirically, the learning process exhibits rapid convergence, with agents typically reaching stable configurations within the first tens of interaction steps. This behavior is mainly attributed to the smooth and physically grounded reward signal based on the Hamiltonian variation, which avoids abrupt or sparse feedback and promotes stable temporal-difference updates.

\subsection{Static Agents}
By setting the specified parameters, we obtain Fig. \ref{fig: Environments} for all four strategies discussed. Here, we have $N = 100$, $\rho = 0.51$. This figure presents four graphs corresponding to each strategy in the order they were described. All graphs represent the same time point, specifically one of the last 10 steps selected.

\begin{figure}[t]
	\centering
	\includegraphics[width=0.5\linewidth]{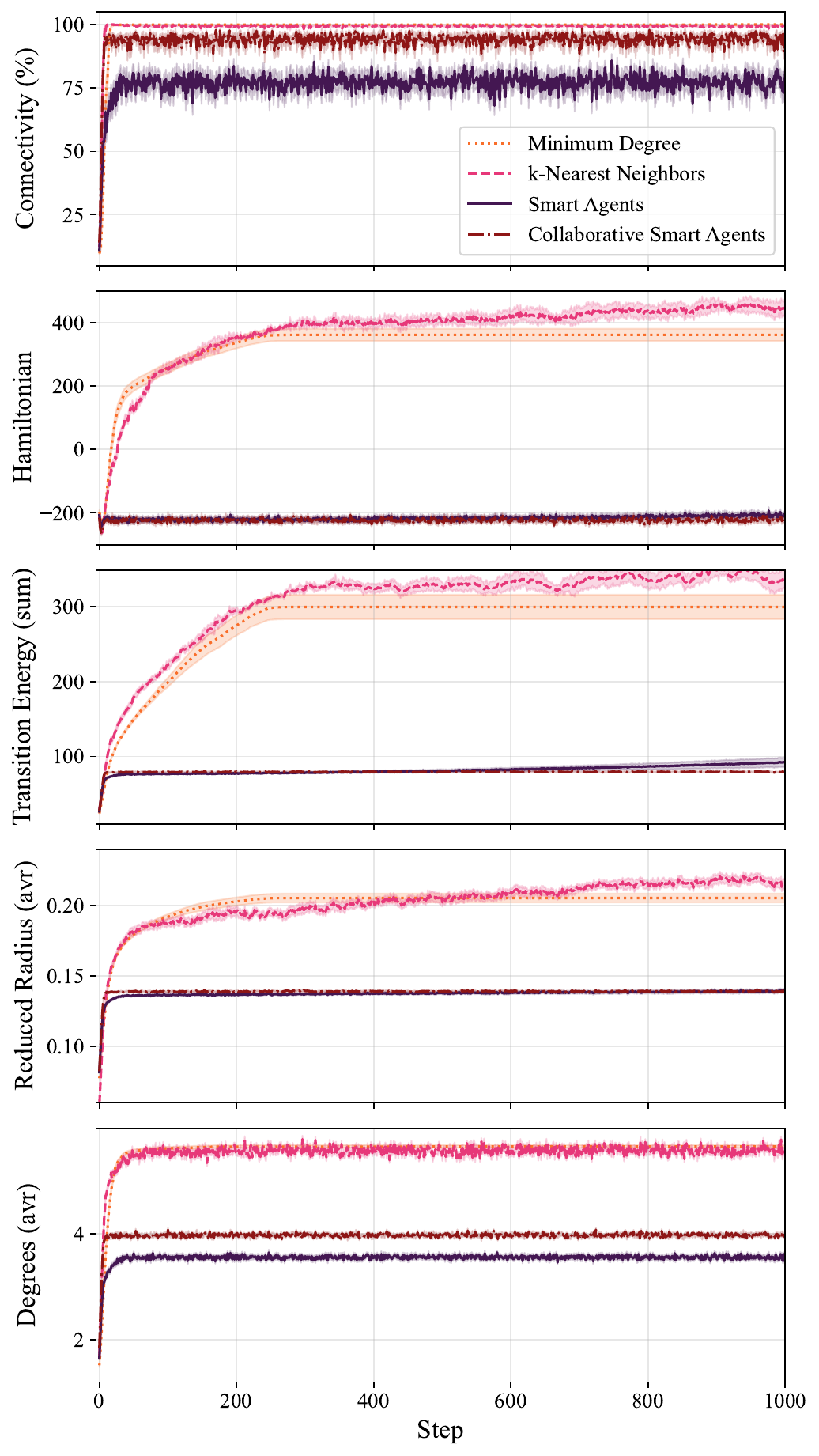}
	\caption{The results of the statistical characteristics of the system for stationary agents are obtained from averaging across twenty different simulation scenarios. In this simulation, $\rho = 0.51$, and the Hamiltonian coefficients are set to $\left[-0.5, 0.1, 0.2, -0.5\right]$. In this series of simulations, the average connection percentage for the minimum 5-degree strategy is $99.8\% \pm 0.05\%$, for the 5-nearest neighbor strategy is $99.4\% \pm 0.04\%$, for the neural network strategy, it is  $77.0\% \pm 0.12\%$, and for the collaboration strategy, it is $94.1\% \pm 0.07\%$	(values are reported as mean $\pm$ SEM over all time steps and simulations).}
	\label{fig: Static}
\end{figure}

The collaboration strategy has outperformed the neural network strategy alone, indicating that even minimal cooperation among neighboring agents can significantly enhance communication across the entire network and increase the connectivity percentage. Additionally, it is observed that the classical strategies, minimum degree and nearest neighbor, have both achieved full connectivity, which was expected since we compelled the agents to connect with a large number of other agents. However, the high connectivity seen in these graphs should not be interpreted as superiority; the costs associated with this state (in terms of transmission radius and energy consumption) are quite substantial. This issue can be better analyzed through the graphs presented in Fig. \ref{fig: Static}. In this figure, the statistical characteristics of the system for all four strategies are illustrated, representing the overall network connection percentage, the total Hamiltonian of the agents according to Eq. (\ref{eq: Hamiltonian}), the total transmission energy consumption $\left( \alpha_{3} \sum r^{2} \right)$, the reduced average transmission radius of the agents $\left( 1/N \sum r/L \right)$, and the average number of their connections $\left( 1/N \sum k \right)$.

By examining the graphs in Fig. \ref{fig: Static}, which represent the average results from an ensemble across twenty different configurations of agent placements in the environment, it is evident that the collaboration graph displays significantly better performance. This strategy not only maintains lower energy consumption but also achieves a higher connection percentage compared to the neural network state. Although the average radii are nearly equal, the collaboration strategy establishes more links, indicating that the agents' radii are adjusted more intelligently to control energy consumption while enhancing connection stability. Additionally, it can be observed how using learning-based models effectively reduces both Hamiltonian values and energy consumption compared to classical strategies.

\subsection{Moving Agents}
We incorporated mobility into the agents to make the simulation more closely resemble real-world models. The movement followed by each agent is considered non-Markovian, meaning that while each agent exhibits random movement, it also tends to move in a specific direction that is randomly assigned and independent of the movement of other agents. Additionally, the boundary conditions we have implemented are reflective; when an agent reaches the boundary, it reflects off that surface like light, maintaining its angle of incidence.

\begin{figure}[t]
	\centering
	\includegraphics[width=0.5\linewidth]{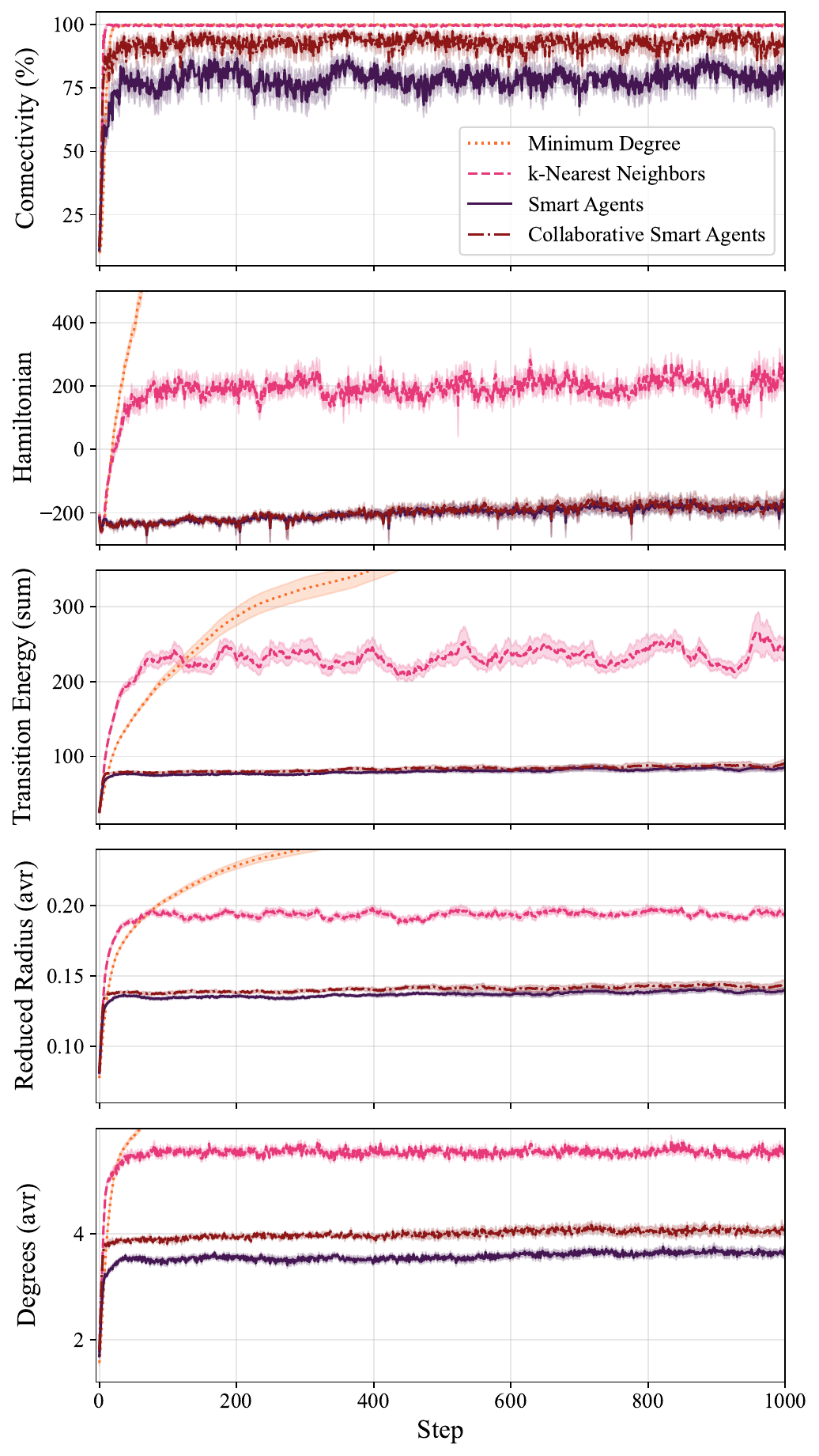}
	\caption{The results of the statistical characteristics of the dynamic system are obtained from averaging across twenty different simulation scenarios. In this simulation, $\rho = 0.51$, and the Hamiltonian coefficients are set to $\left[-0.5, 0.1, 0.2, -0.5 \right]$. In this series of simulations, the average connection percentage for the minimum 5-degree strategy is $99.8\% \pm 0.06\%$, for the 5-nearest neighbor strategy is $99.5\% \pm 0.04\%$, for the neural network strategy, it is  $78.4\% \pm 0.13\%$, and for the collaboration strategy, it is $92.7\% \pm 0.09\%$ (values are reported as mean $\pm$ SEM over all time steps and simulations)}
	\label{fig: Moving}
\end{figure}

The statistical parameters of this series of simulations are illustrated in Fig. \ref{fig: Moving}. As can be seen, the various strategies show little change in the dynamic state of the agents, with the collaboration strategy continuing to outperform the other three. One notable difference observed here is that the fluctuations in the network connection percentage have increased. This is due to the momentary changes in the system, where agents may make suitable decisions for that particular moment, but as other agents move, their connections may be disrupted, forcing them to adapt to the new situation. Additionally, these movements can lead to the temporary formation of one or more clusters of agents that become disconnected from the larger network.

To further make the system behavior more similar to reality, all future simulations are considered as moving agents. Also, given that the collaboration strategy demonstrated better performance across all future simulations, the subsequent results will focus solely on this strategy, omitting the examination of the other methods as unnecessary.

\subsection{\label{sec: hyperparameters Effect}Effect of Some hyperparameters}

\begin{figure}[b]
	\centering
	\includegraphics[width=0.5\linewidth]{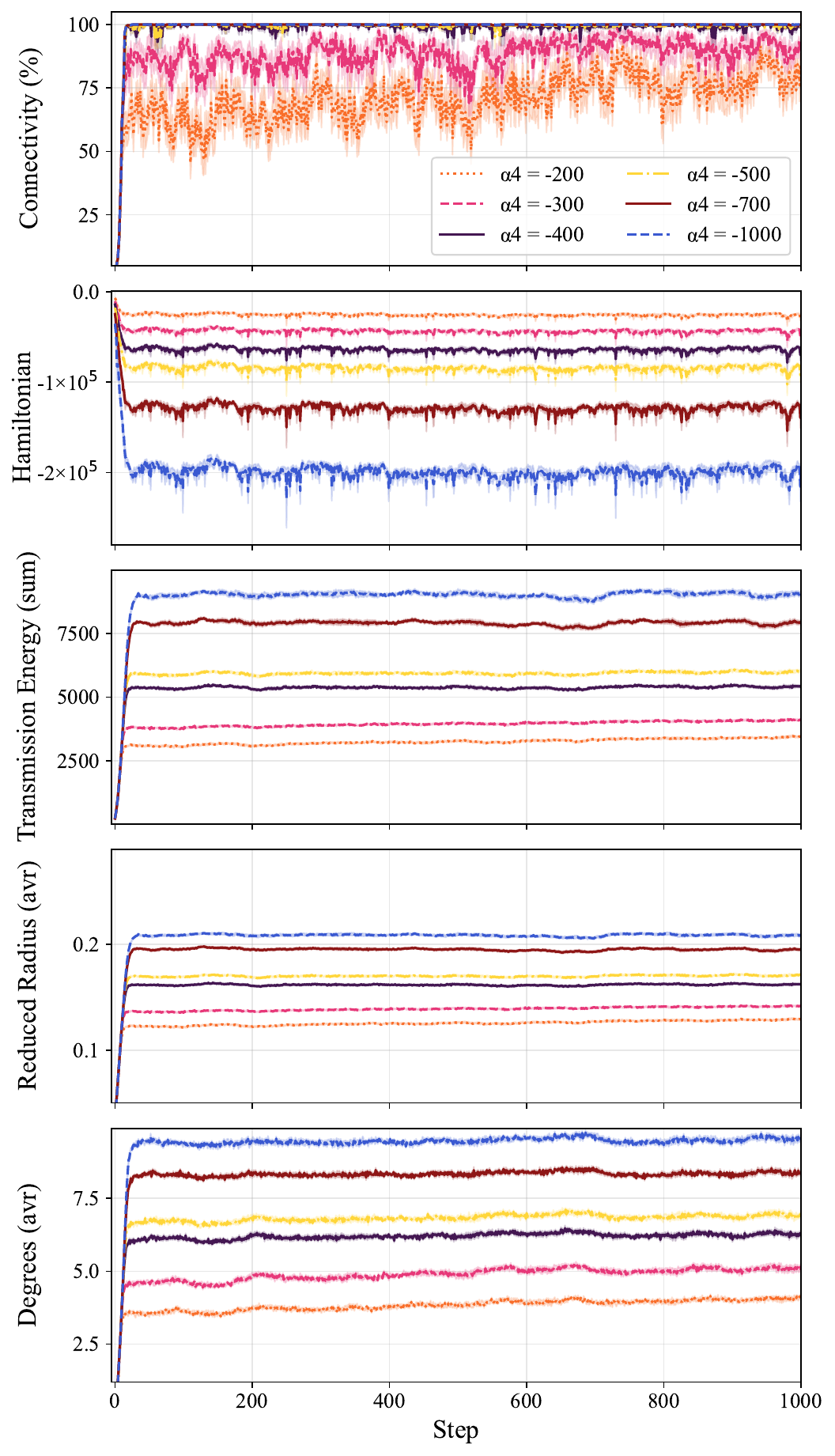}
	\caption{The impact of varying the fourth Hamiltonian term on the overall system parameters. A decrease in the value of $\alpha_{4}$ leads to an increase in the radius, which, due to the constant density of agents, results in a higher average degree of the agents and ultimately enhances the overall connectivity of the system.}
	\label{fig: Statistical Parameters (A Changing)}
\end{figure}

Some of the hyperparameters examined in this system are inherently strongly interdependent, such that changing any one of them can trigger a cascade of effects that significantly influence the overall behavior of the network. These dependencies manifest not only at the local scale (i.e., in the direct interactions between agents) but also in the global structure and dynamics of the entire network. For instance, a variation in one of the coefficients in the system’s Hamiltonian can alter how agents connect, adjust their interaction radius, and even change the stability patterns or transient configurations within the network. Among these hyperparameters, the Hamiltonian coefficients and the particle density of the environment play a central role. These two parameters directly determine the agents’ transmission radius (or, approximately, the number of links each agent maintains with others), and modifying them can drive the network structure from a fully connected state toward less connected or even fragmented configurations.

It was mentioned that the fourth term of the Hamiltonian, by providing a reward, facilitates the creation of long links within the network. We intend to investigate this assertion by varying the $\alpha_{4}$ coefficient.

\begin{figure}[b]
	\centering
	\includegraphics[width=0.5\linewidth]{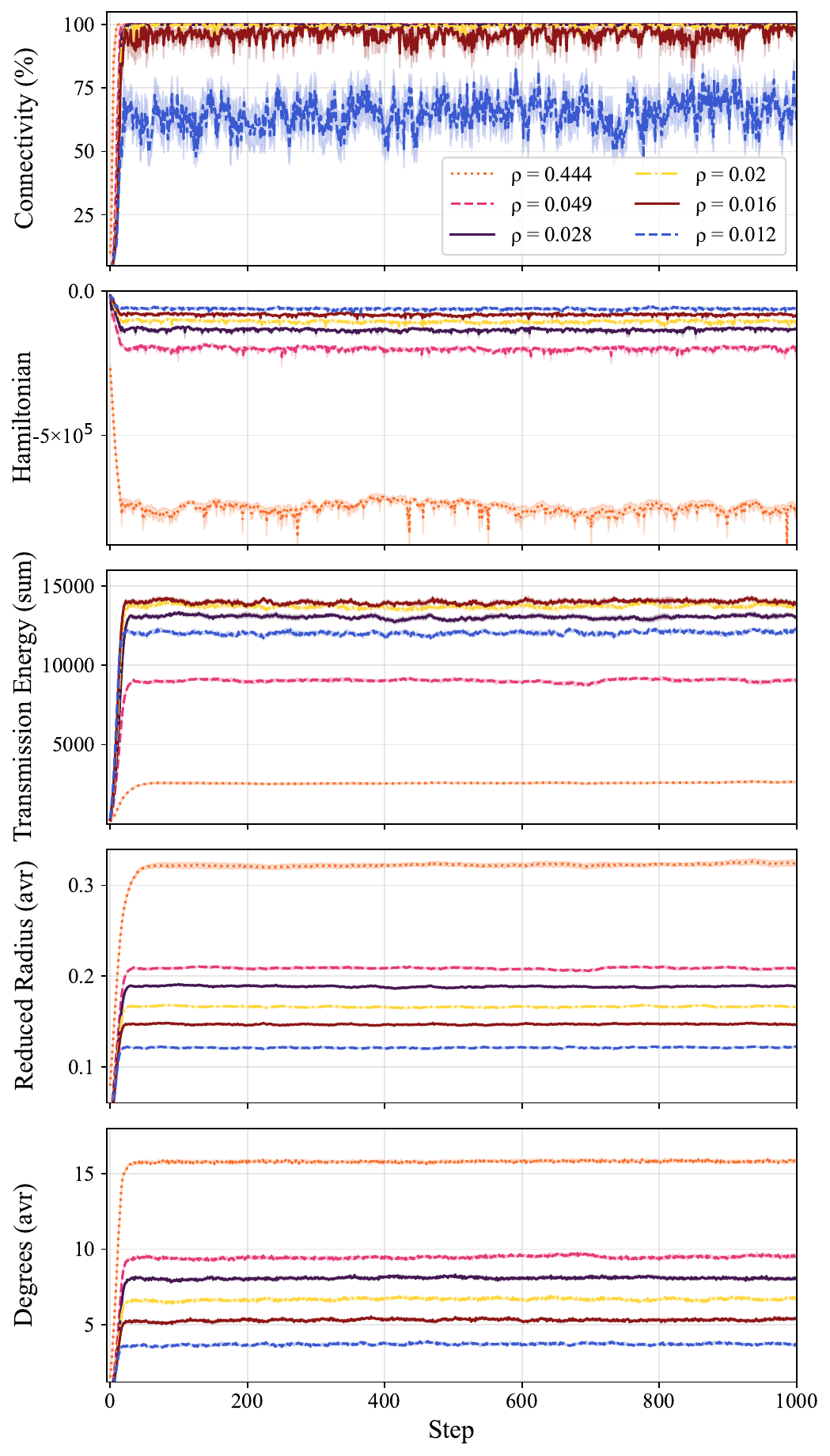}
	\caption{The impact of varying the system's density on the main system variables and the overall behavior of the agents. Increasing the size of the system, and consequently decreasing the density of agents, generally leads to a reduction in the overall connectivity of the network.}
	\label{fig: Statistical Parameters (L Changing)}
\end{figure}

Setting the $\alpha_{4}$ to a more negative value causes an increase in agent connections to further decrease the Hamiltonian, which is interpreted as a reward for the Hamiltonian. As a result, agents are allowed to numerically experience larger radii. This process will continue until the order of the energy-related term, as well as the topological terms in the Hamiltonian, matches the order of the fourth term. In this context, the inhibitory effect in positive terms will depend on the density of agents in the environment: if the density is high, the term $\alpha_{2}k_{i}^{3}$ will dominate, while if the density is low, the term $\alpha_{3}r_{i}^{2}$ will prevail.

Fig. \ref{fig: Statistical Parameters (A Changing)} presents averaged graphs from simulations involving $100$ agents in a space with an approximate density of 0.05, where the coefficient $\alpha_{4}$ takes on different values. As observed, decreasing the value of $\alpha_{4}$ allows agents to be seen at larger average radii, confirming our expectations for the system. Additionally, it can be noted that excessively increasing $\alpha_{4}$ reduces the agents’ radii to the point where network connectivity is no longer complete, resulting in a decrease of up to $70\%$ in connectivity.

Also, we consider different values of particle density and examine the system’s response to these changes through simulations. To conduct this analysis, one can adjust the Hamiltonian coefficients and vary the physical length of the system ($L$), which in effect changes the density, and then observe its impact on the network dynamics. The plots presented in Fig. \ref{fig: Statistical Parameters (L Changing)} show the results for the specific case where the total number of agents is fixed at $N=100$. This constraint ensures that the variations observed arise solely from changes in density (and, consequently, in the mean inter-agent distance), rather than from variations in the total number of agents. This approach allows the effect of density to be assessed in a direct and isolated manner, without interference from other factors.

It can be observed that increasing the system length, or conversely, decreasing the density of agents, can compensate for the overall increase in agent radii, allowing the network to remain connected. However, we witness a phase transition beyond a certain point where the network can no longer maintain connectivity. Additionally, according to the transmission energy graph (which is proportional to the square of the radius), agents will not receive rewards for increasing their radii if they fail to connect with other agents, leading them to stop attempting to expand their radii. In this case, they will settle for a smaller radius that reduces the Hamiltonian cost, resulting in a loss of overall network connectivity. This phenomenon is evident in the graphs for the number density of $0.012$, where the average radius of the agents is $10.8$, which is lower than the number density of the previous case $(\rho = 0.016)$, where the average radius was $11.7$. Furthermore, it is noted that the overall connectivity of the system has decreased by up to $65\%$ due to this smaller radius.

\subsection{Robustness Against Attack and Failure}

\begin{figure}[t]
	\centering
	\includegraphics[width=0.5\linewidth]{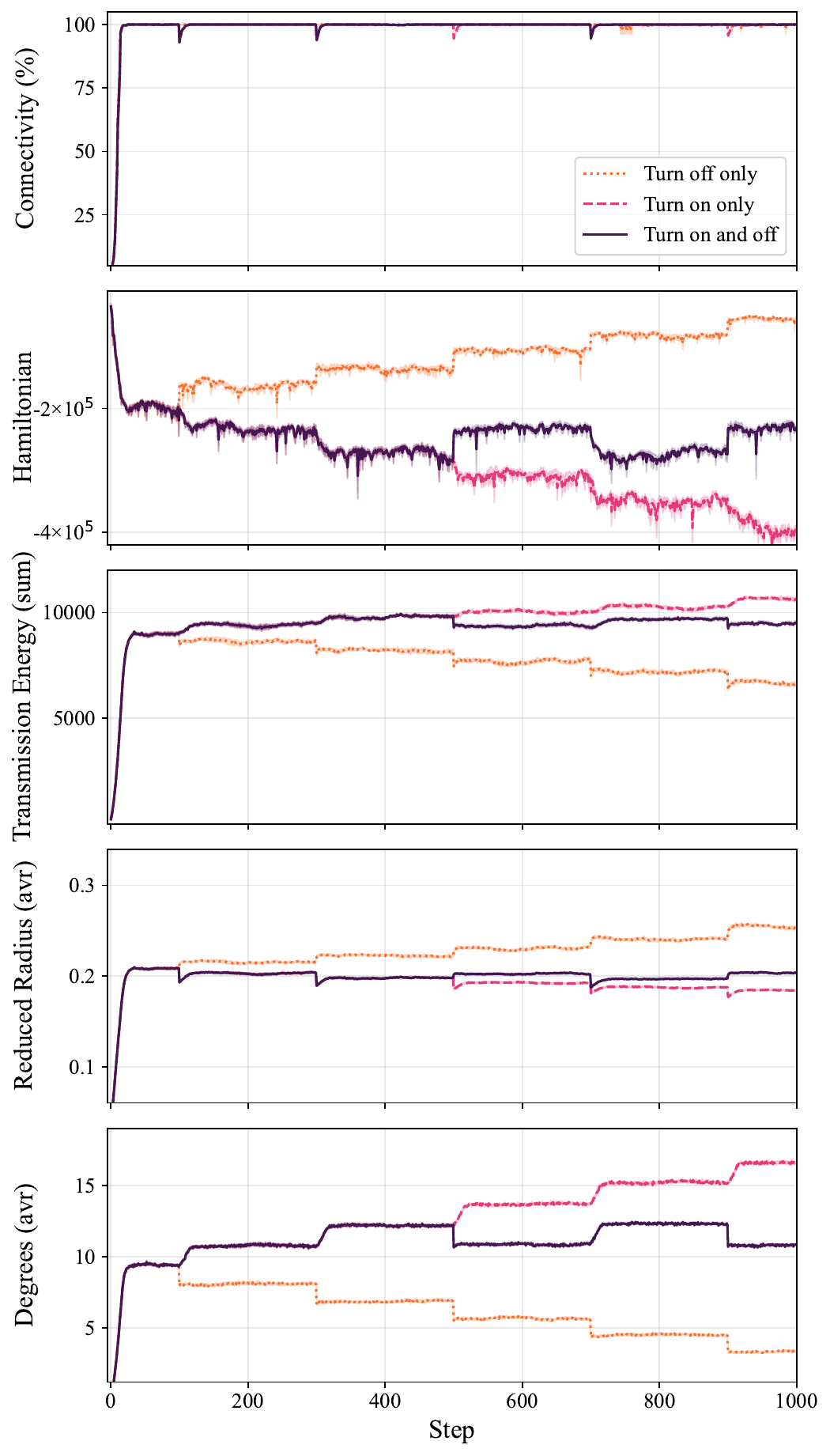}
	\caption{The results of the statistical parameters of the system in the collaboration approach are presented for the removal or addition of $10$ agents every $200$ steps. In this series of simulations, $100$ agents are placed in a space of length $45$, with the Hamiltonian coefficients set to $[-0.5, 0.3, 1.0, -1000]$.}
	\label{fig: Statistical parameters of the On-Off system}
\end{figure}

Another aspect under investigation is the stability and capability of the system to recover and change the types of connections when agents are removed or when new agents are introduced into the environment. To this end, we consider three scenarios: one in which agents are only removed from the environment, another in which agents are only added, and a third in which agents are randomly added or removed. Fig. \ref{fig: Statistical parameters of the On-Off system} illustrates the graphs for these three scenarios, where changes occur every 200 steps, with exactly 10 agents being turned off or on during each iteration.

According to the connection percentage graph, as soon as there is a change in the environment, the network reconstructs itself after a few steps to achieve full connectivity. When agents are added to the system, the average radius decreases, allowing each agent to consume less energy while connecting to a greater number of agents due to the increased density. Conversely, when agents are removed from the environment, the average radius quickly increases, enabling the network to maintain a sufficient average degree despite the higher energy costs associated with this larger radius, thus ensuring complete connectivity.

\begin{figure}[t]
	\centering
	\includegraphics[width=0.5\linewidth]{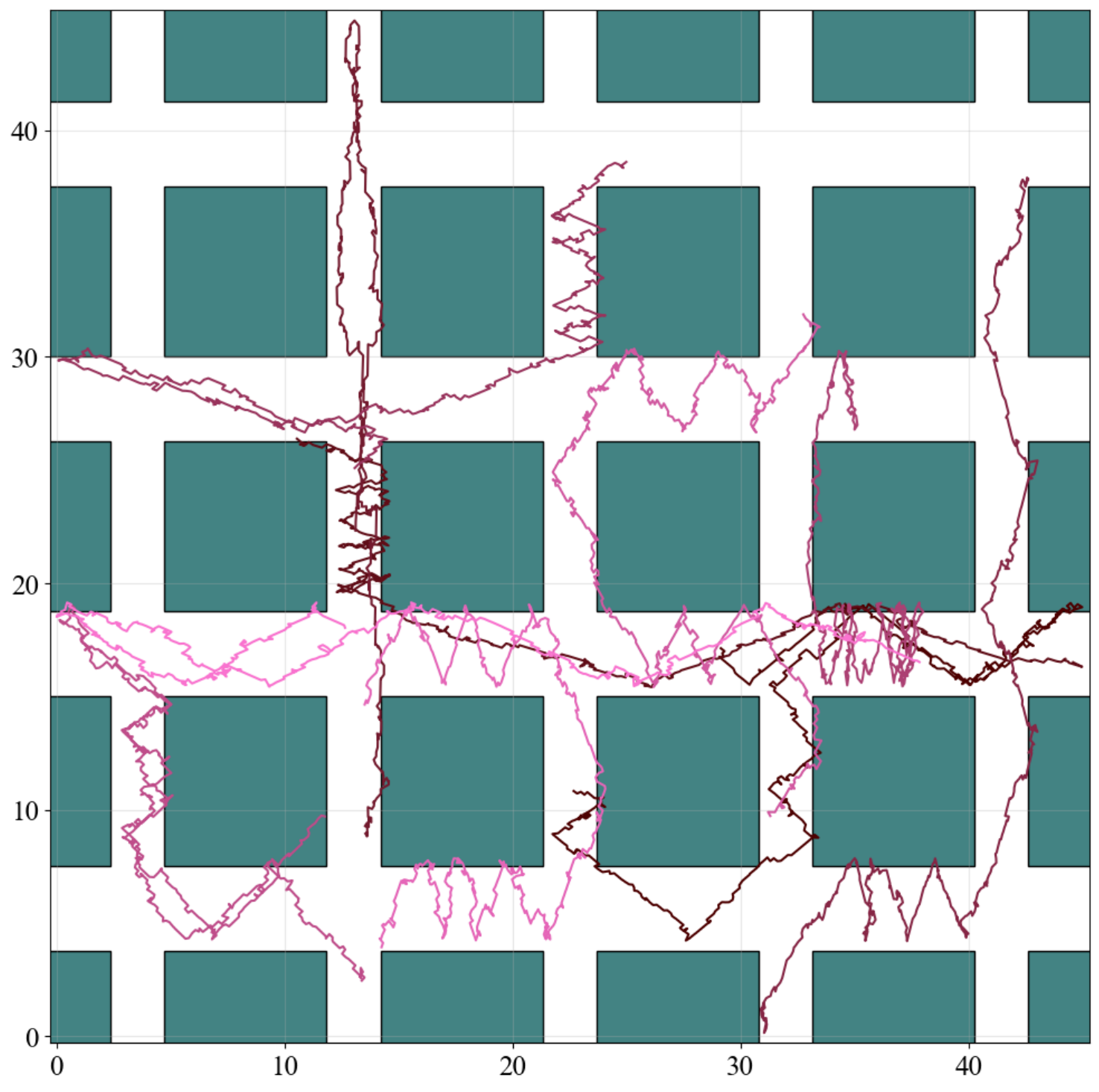}
	\caption{The presence of obstacles arranged in a regular pattern within the simulation environment and the movement of agents among them is depicted in this model, known as the Manhattan model. In this framework, obstacles are treated as building blocks, and agents can only navigate between them, much like moving through streets.}
	\label{fig: Regular Buildings}
\end{figure}
\begin{figure}
	\centering
	\includegraphics[width=0.85\linewidth]{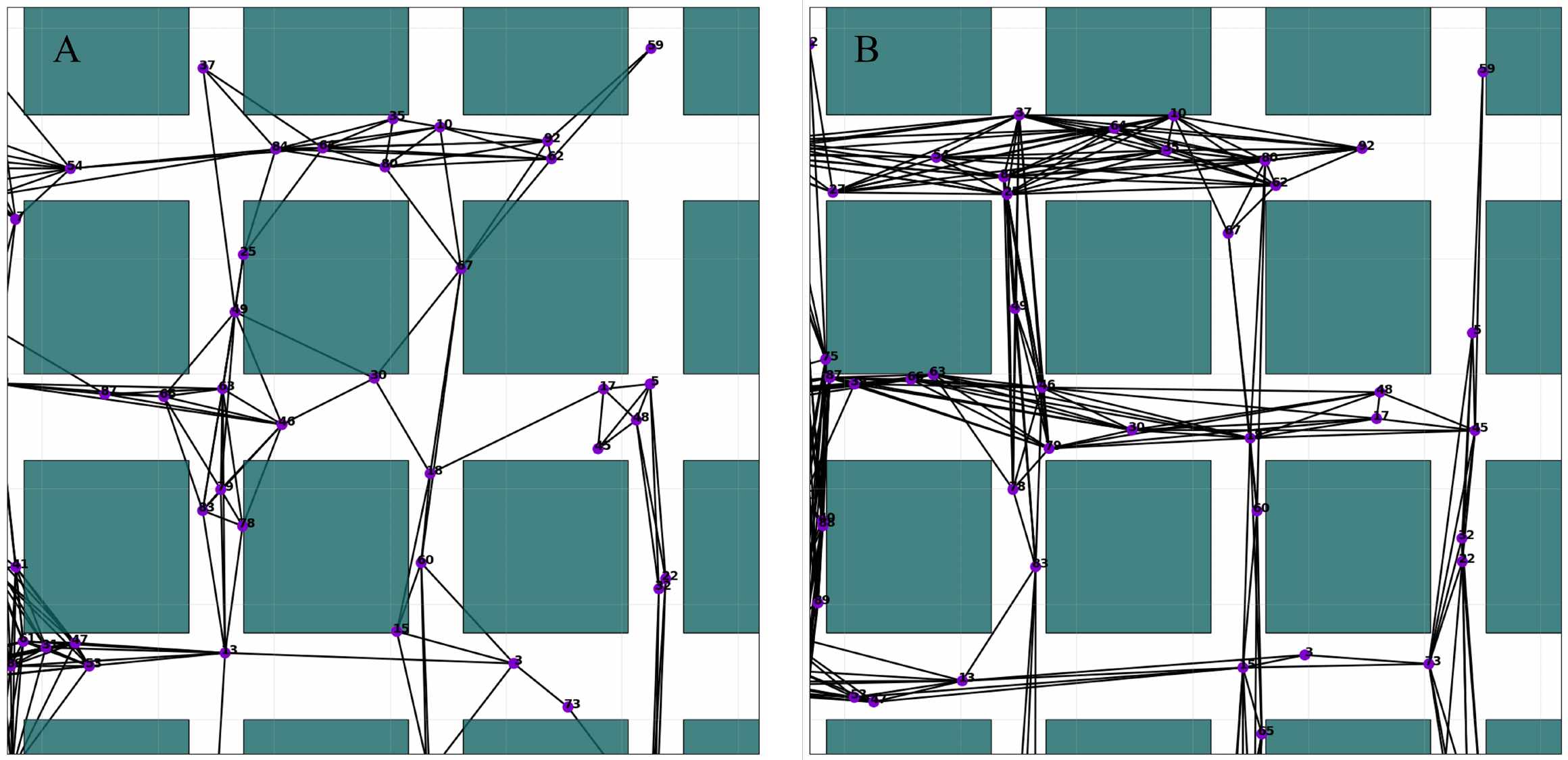}
	\caption{The types of connections between agents in a section of the environment for A) 50\% absorbing obstacles and B) fully wave-absorbing obstacles. This figure is a zoomed-in view of the main environment; therefore, some links extend beyond the image, indicating that they are connected to other agents not shown in this figure.}
	\label{fig: Buildings Zoom}
\end{figure}

According to the presented results, it can be observed that the dynamics of the system can easily accommodate changes in the number of agents while maintaining network connectivity. This connectivity recovery occurs so rapidly that the network deviates from full connectivity for only a few steps. Additionally, the intelligence of the agents allows them to effectively preserve network connectivity even when half of the agents have been removed from the environment, resulting in a density reduction to half. By calculating costs, they manage to maintain a robust connection within the network.

The results discussed above correspond to scenarios in which nodes and links are removed randomly. In contrast, the proposed model demonstrates strong robustness when subjected to targeted removal of high-degree nodes or bridge-like links (red links), as illustrated in Figure \ref{fig: Environments}. Compared with the minimum k-degree approach, the link distribution in our model is more uniform, leading to fewer high-degree nodes. Additionally, the likelihood of the presence of red links is lower than in conventional neural network architectures. Together, these factors contribute to a model that is more resilient to deliberate or organized attacks.

\subsection{Adaptiveness}

Another aspect worth considering is assessing the adaptability of agents to new environments. To this end, we add a number of obstacles to the environment and, while changing the density of agents, examine how their behavior will adapt to achieve the Hamiltonian optimum in the system if a percentage of the transmission radius wave is absorbed or reflected by these obstacles. Therefore, we have introduced regular walls into the environment, similar to buildings (see Fig. \ref{fig: Regular Buildings}). We will examine three scenarios in the simulations: first, where 100\% of the wave passes through the obstacles; second, where 50\% of the wave passes through; and third, where no wave passes through the obstacles at all. The concept of 50\% wave transmission is considered in its simplest form, such that if there is an obstacle between two agents, their visibility radius (transmission range) relative to each other will be half that of the obstacle-free state. A portion of the environment and the type of connections between its agents for two cases -fully absorbing obstacles and 50\% absorption- can be observed in Fig. \ref{fig: Buildings Zoom}.

\begin{figure}[t]
	\centering
	\includegraphics[width=0.5\linewidth]{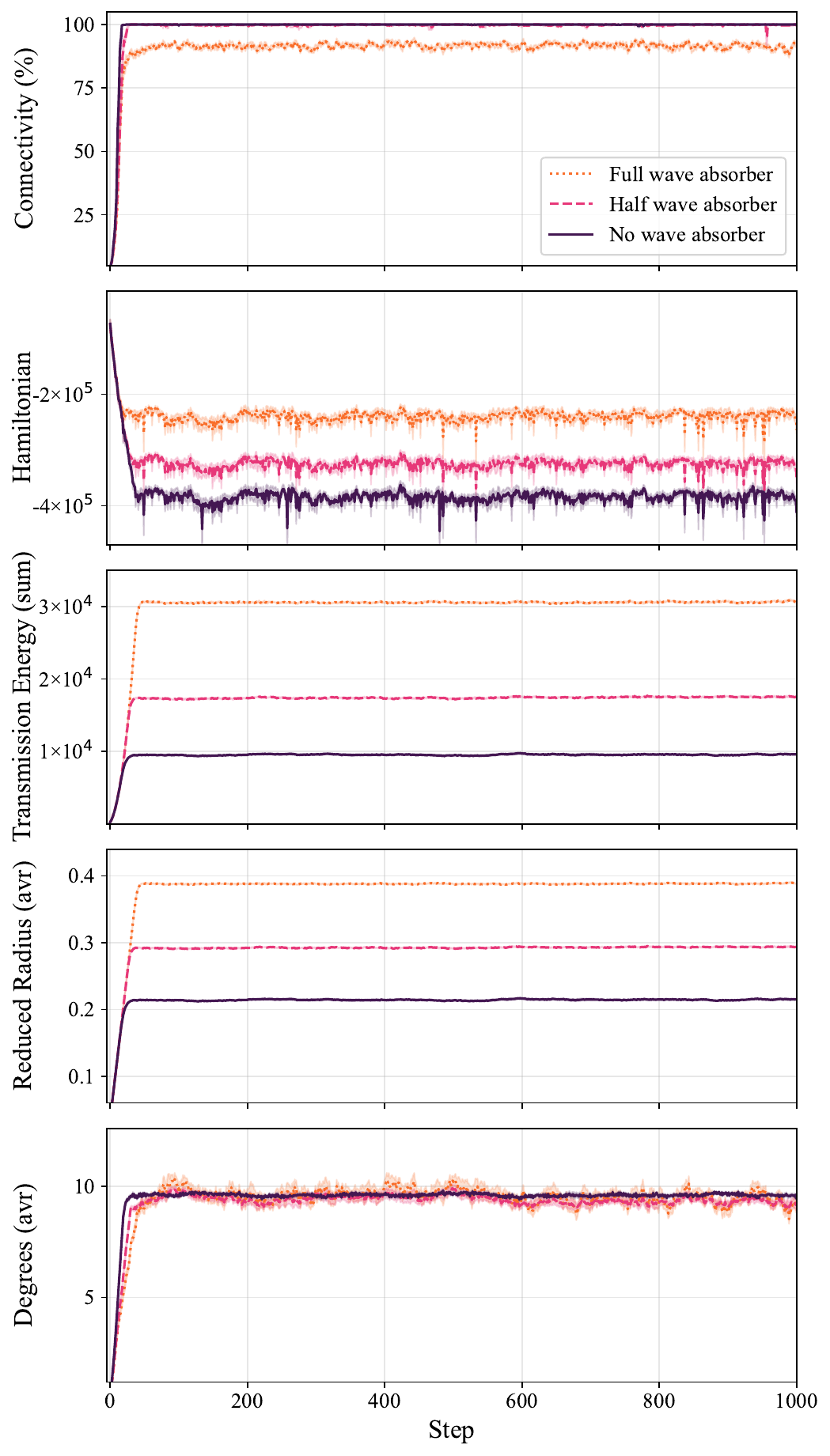}
	\caption{Statistical parameters for the system with obstacles are presented. In this series of simulations, $100$ agents with a number density of $0.05$, and their Hamiltonian coefficients are set to $[-0.5, 0.3, 1.0, -1000]$.}
	\label{fig: Statistical parameters of the buildings system}
\end{figure}

Although the agents were initially trained for obstacle-free environments, it is observed that they quickly learn the new environment \textit{without additional initial training}. They adjust their decisions to maintain the overall network connectivity at the highest possible level using different radii. Initially, the agents had no information about the presence of obstacles in the environment (in the initial weights stored for the neural network). However, by exploring the environment, they learn how to make new decisions to maintain connectivity and optimize their Hamiltonian. This adaptability clearly demonstrates the advantage of the implemented learning method.

Based on the graphs in Fig. \ref{fig: Statistical parameters of the buildings system}, which show the statistical characteristics of these simulations, it can be observed that in all three scenarios, the agents have attempted to maintain a constant average degree. However, this has occurred at different transmission radii depending on their visibility behind the walls. The greater the visibility of an agent from behind the walls, the smaller its radius, resulting in lower energy consumption. Conversely, as visibility decreases, the radius must increase to connect with more agents. This approach increases the likelihood of connecting to agents in key locations (such as intersections), thereby enhancing overall network connectivity.

According to the obtained results, the approach used is effective even in the presence of a large number of obstacles in the environment, and we observe high connectivity. In the case of 100\% wave transmission, which is similar to the normal movement in the environment previously mentioned, we have complete connectivity as before. Additionally, in the 50\% transmission case, we can say with good approximation that we observe complete connectivity. However, in the case where the walls are completely absorbing, the connectivity has decreased to 90\%, which is also expected; this is because at different moments, several agents may be positioned in a corner that remains hidden from the rest, or a situation may arise where a small group is behind multiple obstacles, breaking the overall connection. Nevertheless, 90\% connectivity indicates that despite all these difficulties, the model is still performing well.

			
	
	\subsection{Conclusion}
	
	In this study, we conducted a simulation for self-organizing distributed networks, focusing on the collaboration and reliable communication among the agents in this network to achieve a giant connected network. This network achieves distributed connectivity based on a physical cost function known as the Hamiltonian. In other words, most members of the network create a large main cluster while keeping the defined costs low.
	
	The network's decentralized structure and absence of central control significantly enhance its reliability. In this system, the agents are equipped with artificial intelligence and are released into the environment after a brief initial training. After some time, they effectively learn the environment based on the system's hyperparameters. The result of this learning process is stable connectivity across the entire network while maintaining collaboration between neighboring agents.
	
	The constructed network is robust with respect to the addition and removal of agents due to its overall distributed nature, which is a crucial aspect of network security. Additionally, the network is adaptable to new conditions; even though the agents were initially trained in an environment with a fixed and uniform density, their decisions changed in response to environmental alterations or sudden changes in density, allowing them to maintain overall network connectivity. This demonstrates the advantages of reinforcement learning, where agents are continuously trained and adjust their decisions to align with new conditions. This adaptability can be easily observed in adding obstacles to the environment.
	
	From a computational perspective, each iteration of the algorithm in the proposed simulations involves interactions among all agents to identify neighbors and then make decisions, which results in a time complexity of $O(N^{2})$. In contrast, in the real world the algorithm depends only on local interactions among neighboring agents, and therefore its time complexity is reduced to $O(N⟨k⟩)$, where $⟨k⟩$ denotes the average node degree. In sparse network configurations, this leads to scalability that is nearly linear with respect to the number of agents.
	
	Most importantly, the convergence time of the topology reconstruction process is significantly shorter than the characteristic time scale of agent mobility. Numerical observations show that Hamiltonian‑based learning dynamics reach equilibrium within a limited number of iterations, enabling rapid topology adaptation even in highly dynamic scenarios. To provide a numerical sense of scale, one can say that on a moderately powered computational system, up to $100$ decision‑making steps can be performed per second sufficient time for agents to adapt to network dynamics (for example, in MANETs, VANETs, or any other type of IoT network).


\end{document}